\documentclass[%
 reprint,
 amsmath,amssymb,
 aip,
 jcp,
]{revtex4-1}

\usepackage{amsmath}
\usepackage{graphicx}
\usepackage{dcolumn}
\usepackage{bm}
\usepackage{epsfig,color,xspace,multirow,xr,bbold}
\usepackage[all]{xy}
\usepackage{setspace}
\usepackage{url}

\newcommand{\dGwi}{\Delta G_{\textup{W}\rightarrow\textup{I}}}

\newcommand{\dGwo}{\Delta G_{\textup{W}\rightarrow\textup{O}}}
\newcommand{\dGwol}{\Delta G_{\textup{W}\rightarrow\textup{Ol}}}
\newcommand{\dGim}{\Delta G_{\textup{I}\rightarrow\textup{M}}}
\newcommand{\dGwm}{\Delta G_{\textup{W}\rightarrow\textup{M}}}

\begin{document}

\title{\emph{In silico} screening of drug-membrane thermodynamics
  reveals linear relations between bulk partitioning and the potential
  of mean force}

\author{Roberto Menichetti}
\email{menichetti@mpip-mainz.mpg.de}
\author{Kiran H.~Kanekal}
\author{Kurt Kremer}
\author{Tristan Bereau}
\email{bereau@mpip-mainz.mpg.de}
\affiliation{Max Planck Institute for Polymer Research, 
  Ackermannweg 10, 55128 Mainz, Germany}

\date{\today}

\begin{abstract}
 The partitioning of small molecules in cell membranes---a key
  parameter for pharmaceutical applications---typically relies on
  experimentally-available bulk partitioning coefficients.  Computer
  simulations provide a structural resolution of the insertion
  thermodynamics via the potential of mean force, but require
  significant sampling at the atomistic level.  Here, we introduce
  high-throughput coarse-grained molecular dynamics simulations to
  screen thermodynamic properties.  This application of physics-based
  models in a large-scale study of small molecules establishes linear
  relationships between partitioning coefficients and key features of
  the potential of mean force.  This allows us to predict the
  structure of the insertion from bulk experimental measurements for
  more than 400,000 compounds.  The potential of mean force hereby
  becomes an easily accessible quantity---already recognized for its
  high predictability of certain properties, e.g., passive permeation.
  Further, we demonstrate how coarse graining helps reduce the size of
  chemical space, enabling a hierarchical approach to screening small
  molecules.
\end{abstract}

\maketitle

\section{INTRODUCTION}
\begin{figure}[htbp]
  \begin{center}
    \includegraphics[width=\linewidth]{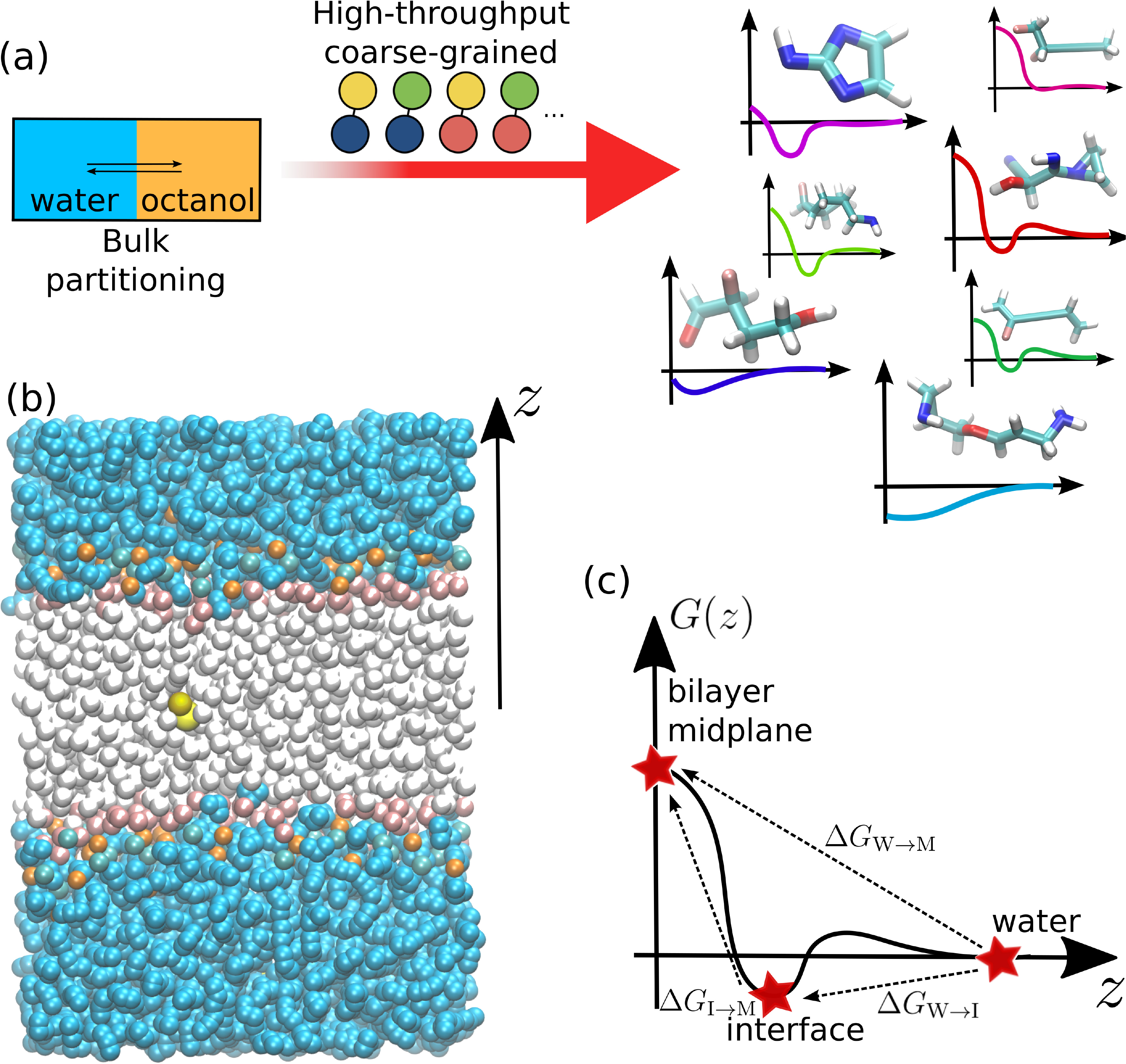}
    \caption{(a) This work establishes linear relationships allowing to predict
      key features of the potential of mean force from a compound's bulk
      water/octanol partitioning coefficient, identified from high-throughput
      coarse-grained simulations.  (b) Simulation setup of a small molecule
      (yellow) partitioning between water (light blue) and the lipid membrane
      (white).  (c) Potential of mean force and the three thermodynamic
      environments of interest: the bilayer midplane, the membrane-water
      interface, and the water phase. }
    \label{fig:intro}
  \end{center}
\end{figure}

The thermodynamic partitioning of small molecules in lipid-bilayer
membrane systems is a key parameter for assessing their suitability
for pharmaceutical applications.\cite{Lopes2016, Shinoda2016,
  Leung2016} It follows that the thermodynamic partitioning is
extremely relevant for high-throughput screening approaches to drug
discovery. Experimentally, for different drug molecules it is
typically obtained by measuring the bulk concentrations of the
molecules in various systems that act as a proxy for the aqueous lipid
membrane systems (e.g., oil/water or Caco-2 cell assays).
\cite{Lopes2016, Breemen2016}  The same property is obtained
\textit{in silico} by simulating the small molecule in an aqueous
environment in the vicinity of a lipid-bilayer membrane
(Fig.~\ref{fig:intro}b).  Further, structural resolution is provided
by the potential of mean force, which describes how the free energy of
the system changes as a function of certain reaction coordinates
(Fig.~\ref{fig:intro}c). In the context of drug-membrane interactions,
the reaction coordinate is often the position of the molecule normal
to the membrane.\cite{Lopes2016,Shinoda2016}  Several studies have
investigated drug-membrane interactions using atomistic molecular
dynamics simulations of specific sets of biomolecules (e.g., amino
acids or drug molecules).\cite{maccallum2008distribution,
  carpenter2014method, lee2016simulation, bennion2017predicting}
Moreover, by means of an integrated measure, atomistic potentials of mean
force were shown to accurately reproduce the experimental transfer free 
energies between water and membrane for 
different compounds.\cite{jakobtorweihen2014predicting}   
  
However, these studies have proven to be severely computationally
expensive, with roughly $10^5$ CPU hours needed to estimate the potential
of mean force for each compound, even with the use of enhanced
sampling techniques (e.g., umbrella sampling).
\cite{neale2011statistical, carpenter2014method}  Given the extensive
computational resources required for these systems, a high-throughput
scheme based on atomistic molecular dynamics simulations is currently
unfeasible for spanning the large regions of chemical compound space
needed to obtain broadly applicable structure-property relationships.

Coarse-grained molecular dynamics simulations provide a means to significantly
reduce the computational expense of fully atomistic simulations while still
capturing the relevant physical properties.\cite{voth2008coarse, Noid2013}
Coarse-grained representations of molecules result from mapping groups of
atoms to coarse-grained ``pseudo-atoms'' or beads. The interaction potentials
of these beads are determined such that the essential properties of the fully
atomistic system are retained at the coarse-grained level. This usually
corresponds to a smoothing of the underlying atomistic free-energy landscape
that preserves its relevant features, allowing for more efficient
sampling. Adapting coarse-grained molecular dynamics simulations to
high-throughput screening of chemical compounds requires flexible and reliable
mapping and force-field parameterization methods.  The coarse-grained Martini
force field provides a robust set of transferable force-field parameters for a
variety of biomolecular systems.\cite{periole2013martini, marrink2013perspective}
 Coarse-grained molecules---akin to a combination
of functional groups---are constructed from a small set of bead types that
encompass a representative spectrum of partitioning free energies between
polar and apolar phases.  We explore a small subset of chemical-compound space
within this model by constructing all possible combinations of one- and two-bead
molecules out of neutral bead types---119 in total.  We then evaluate the
partitioning of each compound in an aqueous lipid-bilayer system from
potentials of mean force and free-energy differences.  Our computational
high-throughput screening scheme establishes linear relationships between key
features of the potential of mean force across compounds, allowing a
semi-quantitative estimation given the Martini representation
of the compound and a \emph{single} parameter:
the water/octanol partitioning coefficient---an experimentally-accessible bulk
property.  Further, the 119 coarse-grained molecules backmap to at least
465,387 different organic small molecules ranging from 30 to 160 Da,
demonstrating a significant reduction in chemical compound space due to the
limited set of bead types of the coarse-grained model.  We thus present two
complementary ways to estimate key features of the potential of mean force
for a large number of small molecules: from coarse-grained simulations and from linear
relationships established in this work.

\section{METHODS}

\subsection{Molecular dynamics simulations}

\label{methods_text}
Molecular Dynamics simulations in this work were performed in GROMACS
4.6.6,\cite{GROMACS-2008} using the Martini force
field.\cite{M-2004,M-2007,M-2008,M-2013} We relied on the standard
force field parameters\cite{de2016martini} with an integration time
step of $\delta t =0.02\hspace{1pt}\tau$, where $\tau$ is the model's
natural unit of time.

A Parrinello-Rahman barostat\cite{parrinello1981polymorphic} and a
stochastic velocity-rescaling thermostat\cite{bussi2007canonical} provided
control over the system pressure ($P=1$ bar) and temperature
($T=300~$K). The corresponding coupling constants were
$\tau_\textup{P}=12\hspace{1.5pt}\tau$ and $\tau_\textup{T}=\tau$.

Bulk simulations consisted of $N_{W}=450$ and $N_{O}=336$ water and
octane molecules, respectively.  A DOPC membrane of $36
\text{ nm}^2$ was generated by means of the INSANE building
tool\cite{wassenaar2015computational} and subsequently minimized,
heated up, and equilibrated. The total number of lipids in the
membrane was $N_{L}=128$ (64 per layer), immersed in $N'_W=1590$ water
molecules.  As usual when using non-polarizable Martini water, we
added an additional $10\%$ of antifreeze particles in the
simulations containing water molecules.\cite{M-2007}

In the case of two-bead molecules, we first considered a representative 
subset of 40 coarse-grained compounds, roughly uniformly covering 
a range of transfer free-energies from water to bilayer midplane of $\dGwm\simeq[-8,14]$ kcal/mol. 
We determined the corresponding potentials of mean force as a function of the
distance $z$ of the compound from the bilayer midplane, $G(z)$, by
means of umbrella-sampling techniques.\cite{torrie1977nonphysical}
We set biasing potentials with a harmonic constant of $k=240$
kcal/mol/$\text{nm}^2$ every $0.1$ nm along the normal to the bilayer
midplane, for a total of 24 simulations. In each of them, two solute
molecules were placed in the membrane in order to increase sampling
and alleviate leaflet area asymmetry.\cite{maccallum2008distribution,
  bereau2014more, jakobtorweihen2014predicting} The total production
time for each umbrella simulation was $1.2\cdot10^5\hspace{1pt}\tau$.
We estimated the free-energy profiles by means of the weighted
histogram analysis
method,\cite{kumar1992weighted,bereau2009optimized,hub2010g_wham} and
the corresponding errors via bootstrapping.\cite{mooney1993bootstrapping}
The same calculations were performed in order to determine the potentials
of mean force for all of the 14 single-bead compounds analyzed in this work.
The computational cost for the reconstruction of each potential of mean force
amounted roughly to $200$ CPU hours. 

In the comparison with atomistic
results\cite{maccallum2008distribution} presented in Fig.~\ref{fig:therm_rel} and Fig.~\ref{fig:pmf}, we
 horizontally shifted the coarse-grained potentials
of mean force by up to $0.4$ nm to correct for discrepancies in the bilayer
thickness between the atomistic and the coarse-grained model, see Ref.~\onlinecite{Bereau2015}.

We herein focus on calculating $\dGwi$ and
$\dGim$, the transfer free-energies between the three different environments---water (W), interface (I), and bilayer 
midplane (M)---along the potential of mean force, see Fig.~\ref{fig:intro}c. In terms of
$G(z)$, these are defined as $\dGwi = G(\bar{z})-G(z\rightarrow\infty)$ and
$\dGim = G(z=0)-G(\bar{z})$, where $\bar{z}\approx 1.8$~nm is the position of
the lipid-water interface with respect to the bilayer midplane ($z=0$).

\begin{figure}[htbp]
  \begin{center}
    \includegraphics[width=\linewidth]{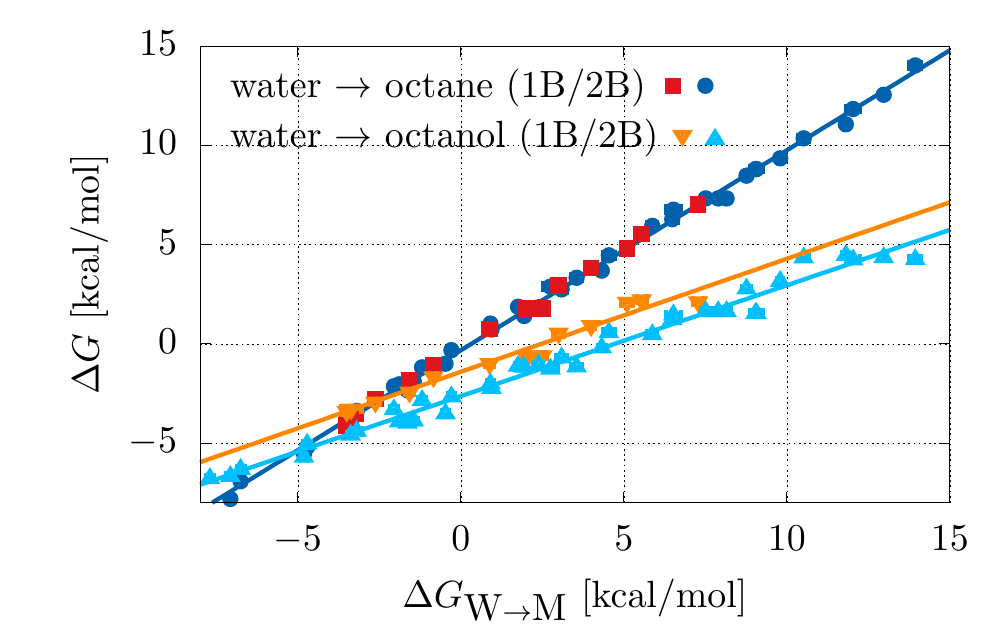}
    \caption{Relationship between the two endpoints of the potential of mean
      force (i.e., $\dGwm = G(z=0) - G(z\rightarrow \infty)$) with the water/octane
      and water/octanol partitioning free energies, in the case of one- (1B) and
      two-bead (2B) compounds. }
    \label{fig:therm_rel_a}
  \end{center}
\end{figure}

The transfer free energies for all 105 coarse-grained two-bead molecules were determined from
alchemical transformations.\cite{chipot2007free} 
Given the excellent agreement between the two end points of a potential of mean force
and the water/octane partitioning (see Sec.~\ref{Linear_rel} and Fig.~\ref{fig:therm_rel_a}), 
as already pointed out in Ref.~\onlinecite{nikolic2011molecular},
the latter was used as a proxy for the hydrophobic core of the membrane.
 
For two compounds
$A$ and $B$, their respective water/interface and interface/octane
transfer free energies are denoted by $\Delta
G^{A}_{\textup{W}\rightarrow\textup{I}}$, $\Delta
G^{B}_{\textup{W}\rightarrow\textup{I}}$, $\Delta
G^{A}_{\textup{I}\rightarrow\textup{O}}$ and $\Delta
G^{B}_{\textup{I}\rightarrow\textup{O}}$.  We can relate these
quantities to alchemical transformations from $A$ to $B$ in fixed
environments via
\begin{equation}
\nonumber
  \Delta G^{B}_{\textup{W}\rightarrow\textup{I}}=\Delta
  G^{A}_{\textup{W}\rightarrow\textup{I}}+(\Delta G^{A\rightarrow
    B}_{\textup{I}}-\Delta G^{A\rightarrow B}_{\textup{W}}),
\end{equation}
\begin{equation}
  \Delta G^{B}_{\textup{I}\rightarrow\textup{O}}=\Delta
  G^{A}_{\textup{I}\rightarrow\textup{O}}+(\Delta G^{A\rightarrow
    B}_{\textup{O}}-\Delta G^{A\rightarrow B}_{\textup{I}}).
\end{equation}

We computed $\Delta G^{A\rightarrow B}_{\textup{I}}$, $\Delta G^{A\rightarrow
  B}_{\textup{W}}$ and $\Delta G^{A\rightarrow B}_{O}$ by means of separate molecular
dynamics simulations at the interface, in bulk water and in bulk
octane. By performing a linear sequence of such transformations covering
all 105 two-bead compounds, we were able
to determine the full set of $\Delta G_{\textup{W}\rightarrow\textup{I}}$ and
$\Delta G_{\textup{O}\rightarrow\textup{I}}$ analyzed in this work.

In the calculation of the $\Delta G^{A\rightarrow B}_{j}$, $j=\text{I,W,O}$, we employed
the multistate Bennett acceptance ratio\cite{shirts2008statistically} (MBAR),
a generalization of the BAR method.\cite{bennett1976efficient} 
MBAR determines the free energy difference $\Delta G^{A\rightarrow
  B}_{\text{j}}$ by appropriately combining the results obtained from
simulations that sample the statistical ensembles generated by a set of
interpolating Hamiltonians $H(\lambda)$, $\lambda\in[0,1]$, with
$H(\lambda=0)=H^{A}$ and $H(\lambda=1)=H^{B}$. Specifically, we made use of
$21$ evenly distributed $\lambda$-points between $0$ and $1$ for each
alchemical transformation and in each environment (interface, water and
octane). The production time for each $\lambda$ point was
$2\cdot10^4\hspace{1pt}\tau$ in bulk water and bulk octane and
$4\cdot10^4\hspace{1pt}\tau$ at the interface.
The cumulative computational cost of performing each alchemical transformation
in water, interface and octane amounted roughly to $60$ CPU hours.

Given the spatial localization of the interface, the free energy $\Delta
G^{A\rightarrow B}_{\text{I}}$ was computed by adding a harmonic potential
between the compound and the bilayer midplane at a distance $\bar{z}=1.8$ nm.

\subsection{Analysis of chemical compound space}

The algorithm developed by Bereau and Kremer\cite{Bereau2015} for the
automated parameterization of the Martini force field was used to coarse grain
approximately 3.5 million small organic compounds, for which 465,387 were mapped to 
Martini molecules consisting of one and two beads. The list of compounds was obtained from the Generated Database
(GDB) of molecules which had up to 10 heavy (excluding Hydrogen)
atoms,\cite{Fink2005,Fink2007} representing most synthetically-feasible small
organic molecules between 30 and 160 Da. The algorithm utilizes a mapping
energy function that is minimized for each molecule so as to optimize both the
number and placement of beads used in its coarse-grained representation. The
bead typing occurs by assigning an existing Martini bead type that has the
best matched water/octanol partition coefficient with that of the molecular
fragment encapsulated by the bead. The partition coefficients of these
fragments are obtained by using ALOGPS,\cite{Tetko2001} a neural-network algorithm that
predicts these values given the chemical structure of the
fragment. The standard mean error associated with this algorithm 
is 0.36 kcal/mol.\cite{Tetko2001, Tetko2002}

\section{RESULTS}

\subsection{Linear relationships}

\label{Linear_rel}

Fig.~\ref{fig:therm_rel_a} shows the
excellent agreement between the two end-points of a potential of mean force
(i.e., $ \dGwm = G(z=0) - G(z\rightarrow \infty)$) and the water/octane partitioning,
$\dGwo$, which illustrates that bulk octane is a good proxy to represent the
hydrophobic core of the bilayer, as already discussed in Ref~\onlinecite{nikolic2011molecular}.
A linear fit for the two quantities provided
\begin{equation}
\label{octane_vs_membrane}
\dGwo=\dGwm -\alpha, 
\end{equation}
$\alpha \approx0.28$, $0.32$ kcal/mol for one-bead and two-bead compounds respectively,
with Pearson correlation coefficients $R^2=0.99$.
As described in the Methods section, this allowed us to determine transfer free energies
with respect to an octane environment, later converting them to the corresponding membrane values.

\begin{figure}[htbp]
  \begin{center}
    \includegraphics[width=\linewidth]{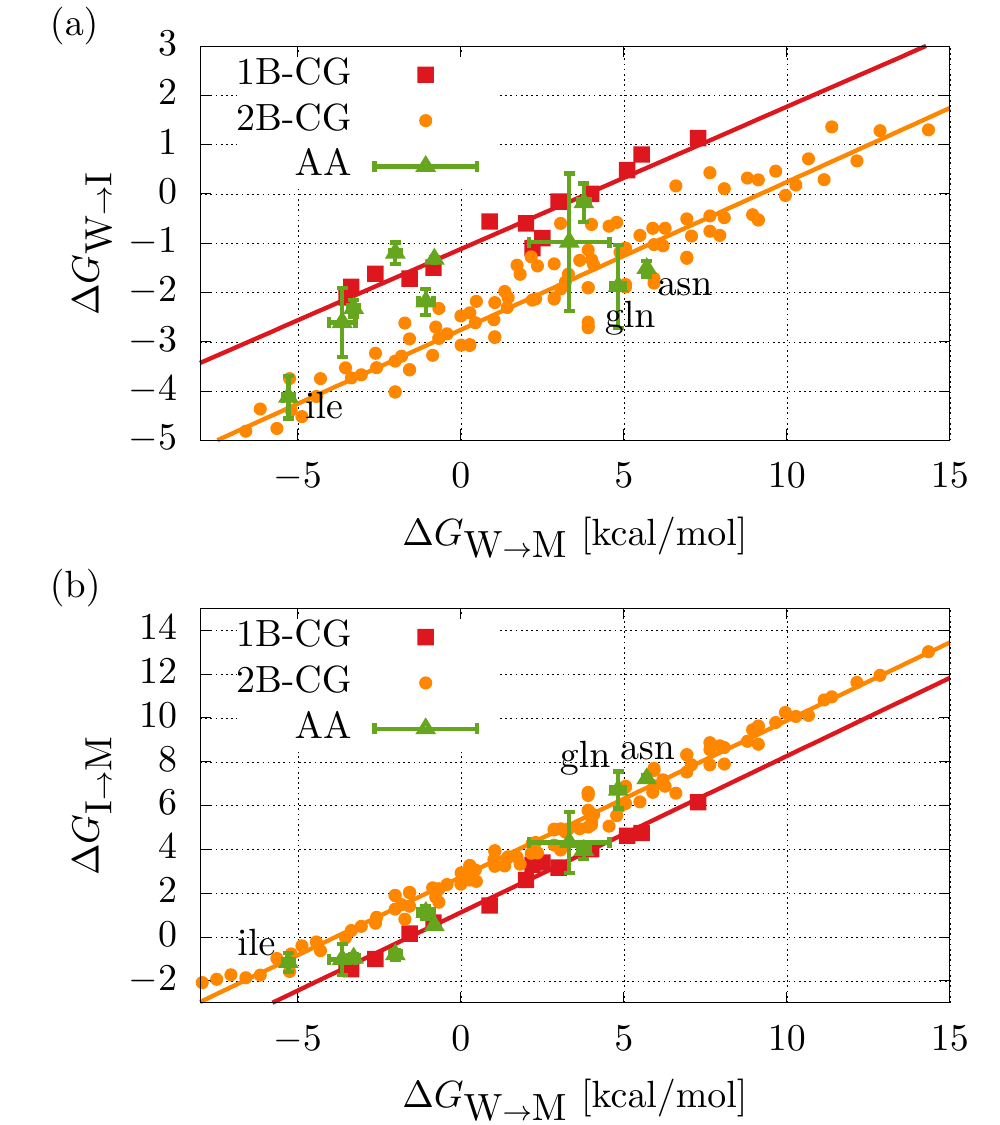}
    \caption{(a) Transfer free energies from water to interface 
      $\dGwi$ as a function of the compound's water/membrane
      partitioning free energy, $\dGwm$.  The red and orange curve correspond to
      coarse-grained estimates for one-bead (1B) and two-bead (2B) molecules,
      respectively. The green points (AA) depict corresponding atomistic
      references of amino-acid side chains.\cite{maccallum2008distribution}
      (b) Transfer free energies from interface to the membrane $\dGim$
      as a function of the compound's water/membrane
      partitioning free energy, $\dGwm$. Color coding follows from (a).
      In both figures, statistically
      significant outliers (see text) are marked with a label (asn, ile, and gln).
      }
    \label{fig:therm_rel}
  \end{center}
\end{figure}

For every compound, the transfer free energies depicted in
Fig.~\ref{fig:intro}c are subject to a thermodynamic cycle that links
the three variables
\begin{equation}
  \dGwi + \dGim - \dGwm = 0.
\end{equation}
Fig.~\ref{fig:therm_rel} illustrates the relationship between these three
transfer free energies for all 119 coarse-grained molecules considered in this
work inserted in a DOPC membrane. In both cases of one- and two-bead 
compounds, beyond the thermodynamic cycle linking
$\dGwi$, $\dGim$ and $\dGwm$, we observe a collapse of the data onto two lines, indicative
of a linear relationship between these transfer free energies.
Moreover, the only difference between one- and two-bead results consists in 
the presence of a simple offset (i.e., same slope) between the profiles.

As a result,
the thermodynamic cycle shown in Fig.~\ref{fig:intro}c 
can be reconstructed from the knowledge of a single variable and the
Martini bead representation of the compound. The error in doing so amounts to
$\approx0.4$.~kcal/mol. These relationships are validated from reference
atomistic simulations of amino-acid side
chains,\cite{maccallum2008distribution}, where we consider only atomistic
compounds whose Martini representation consists of a single bead. 
While most points fall within the linear fit from the
single-bead coarse-grained data, we observe three statistically
significant outliers: asparagine (asn), isoleucine (ile), and glutamine (gln).
These molecules lie on the data corresponding
to two-bead compounds, although their Martini representation consists of 
a single bead.\cite{M-2008} The origin of such discrepancies will be
explained below. The comparison of atomistic and Martini potential
of mean force for protein side-chains was already performed in Ref.\onlinecite{M-2008}.

\begin{figure}[htbp]
  \begin{center}
    \includegraphics[width=\linewidth]{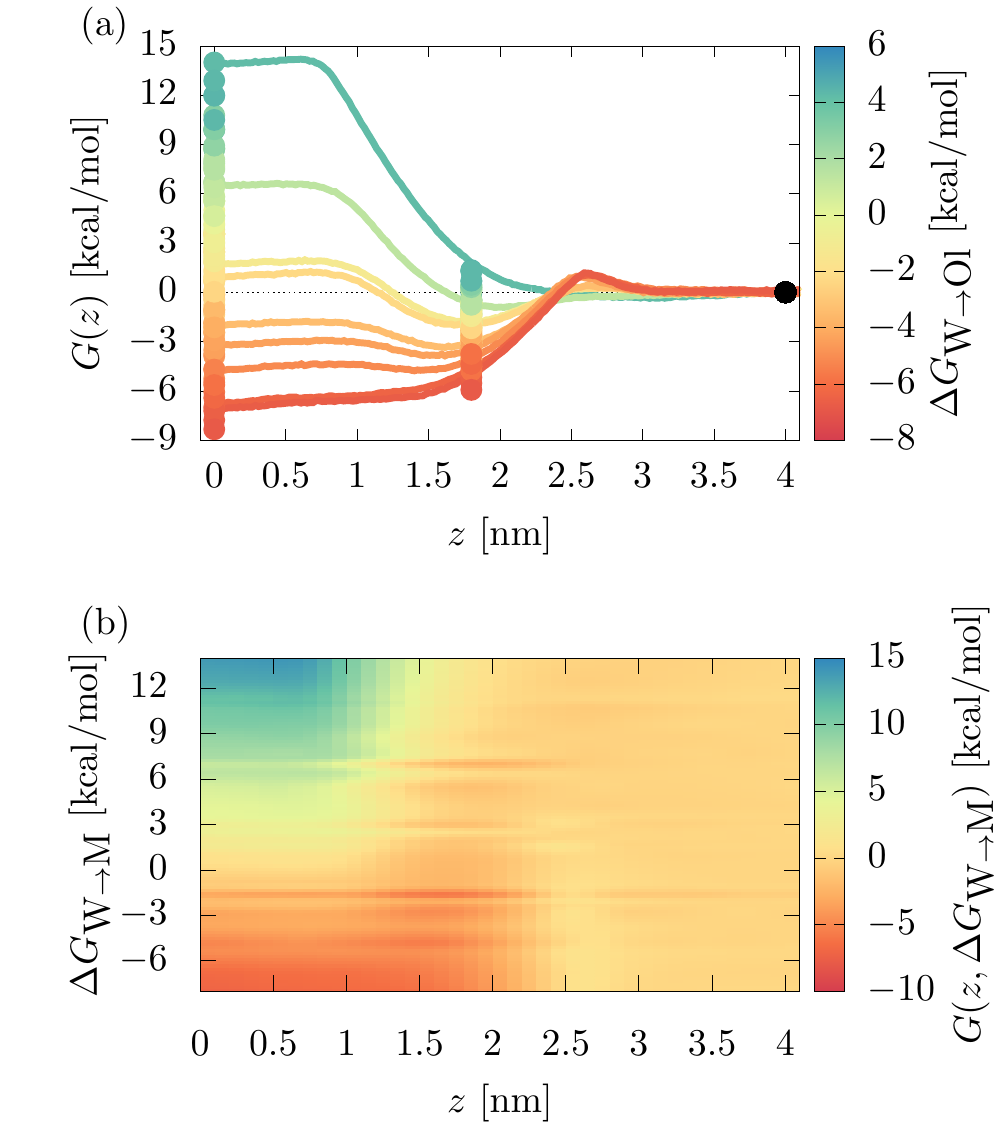}
    \caption{(a) Representative potentials
      of mean force of various Martini compounds as a function of the normal
      distance to the bilayer midplane.  The color range denotes the
      water/octanol partitioning of the small molecule.  Large circles
      correspond to estimates from the thermodynamic relation
      extracted in this work. 
   (b) Two dimensional map of the free energy surface $G(z,\dGwm)$ for
      a small molecule, as a function of its distance from the DOPC bilayer
      midplane $z$ and its membrane/water partitioning free energy $\dGwm$.}
    \label{fig:fe_surface}
  \end{center}
\end{figure}

Remarkably, the relationships between transfer free energies displayed in
Fig.~\ref{fig:therm_rel} can further be linked to a compound's water/octanol
free-energy $\dGwol$, given its accurate linear relation with $\dGwm$, see Fig.~\ref{fig:therm_rel_a}.  
A fit of the data provided
\begin{equation}
\label{w-o/w-ol}
\dGwm =\gamma\dGwol + \delta,
\end{equation}
with $\gamma=1.70,~1.75$ and $\delta=2.51,~4.69$ kcal/mol for one- 
and two-bead compounds, with $R^2=0.97$.
Given a compound's experimentally determined bulk measurement and Martini representation,\cite{Bereau2015}
we can thereby reconstruct the three main points of the potential of mean force,
as shown in Fig.~\ref{fig:fe_surface}a. We rationalize these findings by
noting the suitability of the octanol environment as a proxy for the membrane
interface.  Similarly, we showed the appropriateness of octane for the bilayer
midplane (Fig.~\ref{fig:therm_rel_a}).  Indeed, both water/alcohol and water/alkane coefficients
correlate with blood-brain partitioning.\cite{toulmin2008toward} Therefore,
the relationships in Fig.~\ref{fig:therm_rel} stem directly from the linear
correspondence between water/octane and water/octanol transfer free energies
(which can be deduced from the linear relations shown in Fig.~\ref{fig:therm_rel_a}).
From the model's perspective, the linear relations 
are not entirely unexpected, as Martini describes hydrophobicity by a set of equally-sized 
Lennard-Jones particles, with varying well-depths.
Interestingly, these relationships also hold at the atomistic level.
At infinite dilution, the difference in partitioning of a single
small molecule between water and either octane or octanol is due to a single
hydrogen bond. 
We suspect that, at the atomistic level, 
the impact of this hydrogen bond on the partitioning behavior 
strongly informs the linearity observed, although the exact mechanism
remains unclear.  We further remind the reader that the relationships presented here depend
strongly on the molecular weight (see the differences between one- and two-bead
molecules in Fig.~\ref{fig:therm_rel}).  
Potentials of mean force of larger
compounds\cite{carpenter2014method} do not follow the relationships presented
in Fig.~\ref{fig:therm_rel}.  Whether other relationships can be determined for 
these molecules will be the subject of future work.

The statistical errors displayed by the coarse-grained simulations are
marginal: less than $0.1$~kcal/mol. However, a comparison of experimental measurements of
the water/octanol partitioning free energies of several hundred small molecules against Martini predictions
yielded a mean-absolute error of 0.79~kcal/mol.\cite{Bereau2015} 
Given the relation between the water/octanol and water/midplane curves of
Fig.~\ref{fig:therm_rel_a} we deduce from it a mean absolute error on features of the
potential of mean force of approximately 1.4~kcal/mol.
Further, the error associated with the
fitted lines on Fig.~\ref{fig:therm_rel} amounts to an overall error of
roughly 1.8~kcal/mol in reconstructing the main points of the potential of mean 
force---at the bilayer midplane and at the interface, see circles in Fig.~\ref{fig:fe_surface}a---by 
using as input only the experimental water/octanol partitioning free energy
of a compound.
At the atomistic level, too few potentials of mean force are available to
provide errors across chemical compounds.

The linearity observed between the free-energy
barrier $\Delta G_{\textup{W}\rightarrow\textup{I}}$ (equivalently $\dGim$) and the
water/membrane partitioning free-energy $\dGwm$ suggests the
possibility of looking for an approximately smooth two dimensional
free-energy surface $G(z,\dGwm)$ across chemical space, hence as a
function of $\dGwm$ as well as of the distance from the bilayer
midplane $z$.

In the case of two-bead coarse-grained molecules, we then constructed a two dimensional
map of the free-energy surface $G(z,\dGwm)$ starting from the 
 set of $40$ potentials of mean force that were determined by means of umbrella
 sampling simulations, covering a range $\dGwm\simeq[-8,14]$ kcal/mol. Results are shown in
Fig.~\ref{fig:fe_surface}b.

The correlations shown in Fig.~\ref{fig:therm_rel} between $\Delta
G_{\textup{W}\rightarrow\textup{I}}$ and $\dGwm$ for different
compounds correspond, on this surface, to the set of points
$G($$\bar{z}~=~1.8$ nm$,\dGwm)$.  Apart from minor fluctuations, it is
evident how the overall smoothness of the surface on the lines with
constant $z$ allows us to identify the existence of an average
functional relationship between $\dGwm$ of a compound and its
potential of mean force $G(z)$ for every value of $z$.  As an example,
a small free-energy barrier located at $z\approx2.5$ nm is present for
all the compounds with $\dGwm\in[-8,0]$ kcal/mol. Small shifts in z may result 
from bilayer-thickness discrepancies between atomistic
and coarse-grained simulations.\cite{Bereau2015}

In this work we focused on the reconstruction of key
features of the potential of mean force (i.e., the water/interface and interface/membrane
transfer free energies, $\dGwi$ and $\dGim$).
The results shown in Fig.~\ref{fig:fe_surface}b further suggests that a knowledge
of the water/membrane partitioning free energy of a
compound, which can be obtained from the corresponding water/octanol one via
the linear relation reported in Eq.~\eqref{w-o/w-ol}, allows for a
semi-quantitative reconstruction of the whole potential of mean force $G(z)$.

\begin{figure}[htbp]
  \begin{center}
    \includegraphics[width=\linewidth]{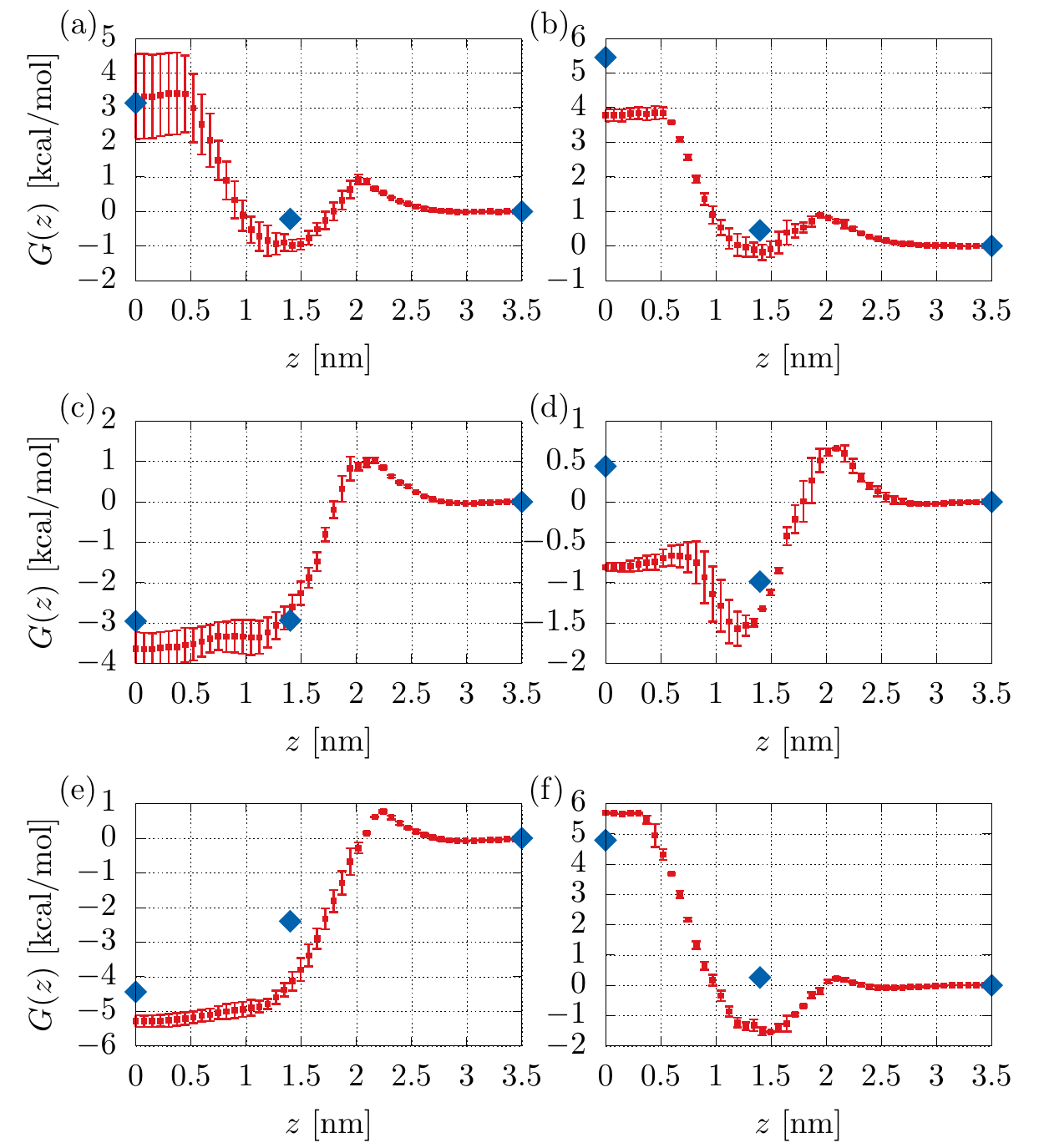}
    \caption{Comparison between the atomistic potentials of mean force of 
    Ref.~\onlinecite{maccallum2008distribution} (red lines) and the predictions obtained
    via the linear relations presented in this work (blue squares), 
    for the insertion of amino-acid side chains in a DOPC membrane: (a) threonine, 
    (b) serine, (c) leucine, (d) cysteine, (e) isoleucine (ile), (f) asparagine (asn).}
    \label{fig:pmf}
  \end{center}
\end{figure}

We now quantitatively test our results by comparing the reconstruction of the
potential of mean force obtained through the linear relationships presented in 
Fig.~\ref{fig:therm_rel} with the atomistic results for amino-acid 
side chains of Ref.~\onlinecite{maccallum2008distribution}, which map to Martini single-bead molecules.
In practice, for each compound we started from the water/octanol 
partitioning free energy obtained by means of the ALOGPS neural network,\cite{Tetko2001} 
which was then converted to a water/membrane free-energy by means of Eq.~\eqref{w-o/w-ol}. 
We then used the linear relations of Fig.~\ref{fig:therm_rel} to determine the water/interface 
and interface/midplane transfer free energies. Results are shown in Fig.~\ref{fig:pmf}.
In the first four panels, corresponding to threonine, serine, leucine and cysteine, our predicted 
free energies are in good agreement with the atomistic results (within our $1.8$ kcal/mol error estimate). 
Therefore, our method based on coarse-grained simulations is capable of providing a good approximation 
to key features of the atomistic potential of mean force. On the other hand, by means of an integrated measure on
the potential of mean force, atomistic results were shown to accurately 
reproduce the experimental transfer free energy
between water and membrane for different compounds.\cite{jakobtorweihen2014predicting} 
The agreement between coarse-grained predictions and atomistic/experimental
results has been already described in detail in previous works.\cite{M-2008,Bereau2015}
These results further suggest the important role of coarse-grained
models as a reasonably predictive tool to capture the fundamental
properties of biological soft matter.
In panels (e) and (f) of Fig.~\ref{fig:pmf} we show the case of isoleucine (ile) and asparagine (asn), 
two of the three statistically significant outliers observed in the linear relations 
presented in Fig.~\ref{fig:therm_rel}. As a consequence, the predicted free energies 
do not agree with the atomistic results within our error estimate.
The origin of these discrepancies is very different in the two cases,
and deserves to be discussed separately.
The water/membrane free energy of isoleucine lies outside the interval covered by the Martini single-bead
compounds, i.e., outside the range of free energies in which
linearity is observed, due to the limited number of Martini beads (see Fig.~\ref{fig:therm_rel}). 
The introduction of more apolar beads in Martini could solve this issue, and work in this direction is in progress.
In the case of asparagine and glutamine, which display similar potentials of mean force, the water/interface 
(and consequently interface/membrane) free-energy is underestimated if one uses the linear relation
observed for single-bead compounds  (see Fig.~\ref{fig:therm_rel} and Fig.~\ref{fig:pmf}, panel (f)).
More interestingly, the datapoints corresponding to these molecules are in agreement with
those of two-bead compounds. 
As both molecules are composed by a combination of a polar and an apolar group, this suggests
that their representation in terms of spherical single-bead molecules is incapable of capturing
their amphipathic nature, which in turn generates an underestimation of the predicted free-energy barriers.
Therefore, the natural coarse-grained representation of both compounds
would be that of a two-bead molecule, as Fig.~\ref{fig:therm_rel} suggests.

\begin{figure}[htbp]
  \begin{center}
    \includegraphics[width=\linewidth]{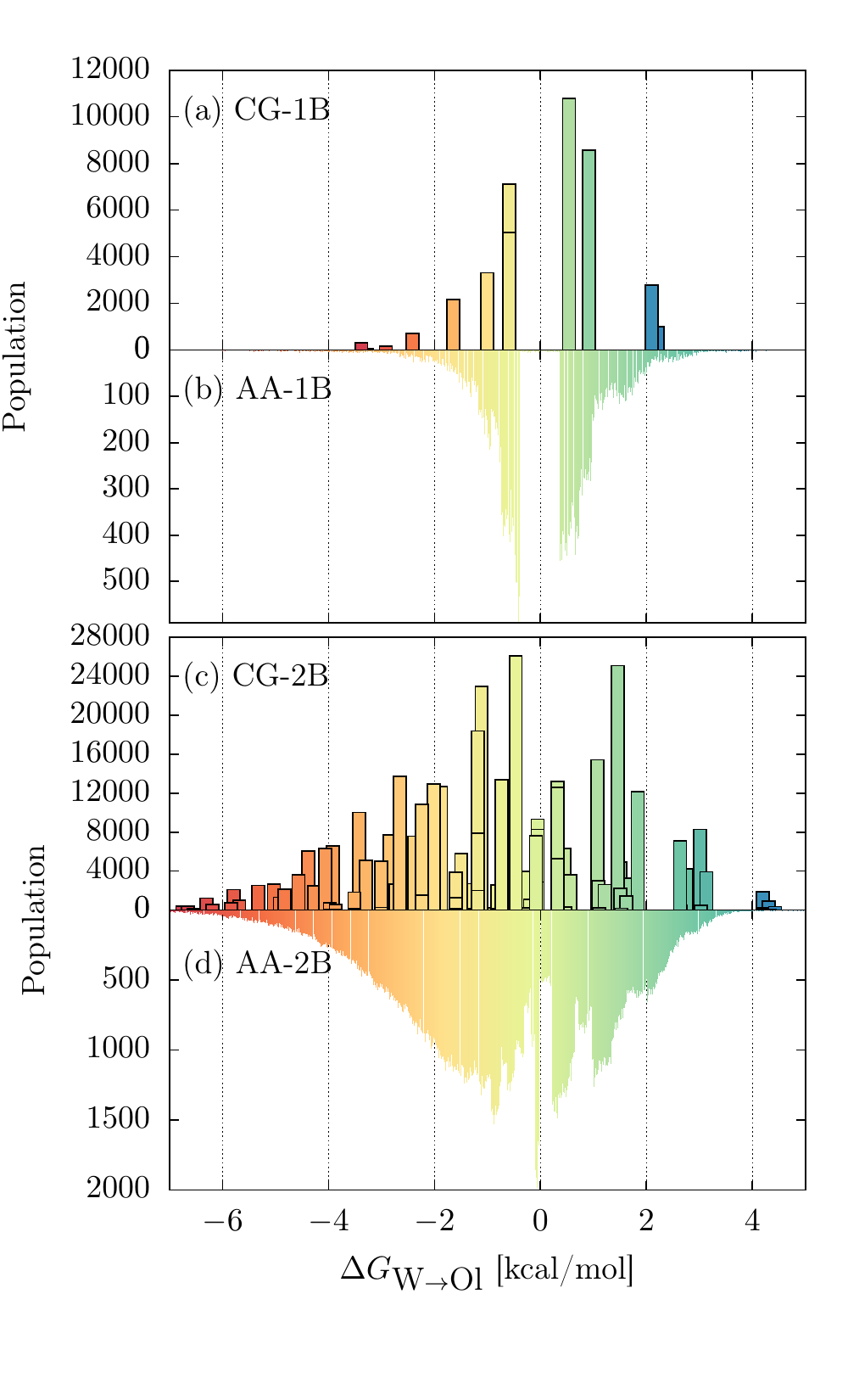}
    \caption{Histograms of 465,387 small molecules extracted from GDB that map
      onto one-bead or two-bead coarse-grained representations.  (a),(c) Coarse-grained and (b),(d) atomistic
      populations as a function of water/octanol partitioning free energy. The width of 
      the bars in (a),(c) have no physical significance and are simply for the 
      reader's convenience.}
    \label{fig:pop}
  \end{center}
\end{figure}

\subsection{Coarse-graining reduces the size of chemical compound space}

\label{Cg_compounds}

Alternatively to the method presented in Sec.~\ref{Linear_rel}, we identify an 
independent means to estimate the potential of
mean force.  We recently showed how the Martini model groups molecules into
fewer coarse-grained representations, thereby effectively reducing the size of
chemical compound space.\cite{Bereau2015}  This grouping stems from the
discrete set of bead types of the Martini model, which assigns the same representation
to groups that are chemically similar.  To estimate this grouping, we have coarse-grained compounds
 from the Generated Database\cite{Fink2007} of molecules up to ten heavy atoms.  
In Fig.~\ref{fig:pop}, we show the distributions of compounds that map to any one 
of the one- and two-bead coarse-grained representations considered here, as a function of the water/octanol
partitioning. The atomistic distributions of Fig.~\ref{fig:pop}b,d were obtained using 
the ALOGPS neural network.\cite{Tetko2001} Despite the uneven spacing of the water/octanol partitioning 
free energies of the coarse-grained molecules, the atomistic distributions are roughly reproduced 
by the coarse-grained distributions in Fig.~\ref{fig:pop}a,c, except for small artifacts in the strongly 
polar regime (i.e., $\dGwol \gtrsim 2.0$~kcal/mol). In total, we identified 465,387 unique 
molecules, representing most synthetically-feasible small organic molecules between 
30 and 160 Da. This many-to-few mapping arises solely from the limited representability of
thermodynamic properties of chemical groups, rather than the coarser
structural representation (i.e., atoms to beads). The removal of chemically and structurally specific information
present in atomistic simulations is traditionally viewed as a necessary drawback for access to 
otherwise computationally prohibitive simulations. However, it is precisely this drawback 
that enables a single coarse-grained simulation to be representative of a large number 
of small molecules, as degenerate chemical groups are mapped to the same bead type. As a 
counterexample, fixing a Martini-like mapping in combination with a non-transferable, chemically-specific 
parametrization (e.g., as in most bottom-up, structure-based models) would prevent any reduction in chemical space. 
This work thereby introduces the ability for transferable coarse-grained models to screen 
large numbers of small molecules.  

\subsection{Permeability coefficient}

Beyond the structural resolution of a large number of compounds across
a membrane interface, the present results provide valuable information
for other properties of interest.  For instance, the permeability
coefficient,\cite{diamond1974interpretation, marrink1994simulation} 
$\log P$---a measure for passive permeation that relies on the
potential of mean force---has been shown to better correlate with
blood-brain barrier permeability compared to water/octanol
partitioning (correlation of 0.7 and 0.9, respectively).\cite{carpenter2014method}  
In the framework of the inhomogeneous solubility-diffusion
model,\cite{diamond1974interpretation,marrink1994simulation}  
the permeability coefficient $\log P$ of a compound is defined as\cite{lee2016simulation}
 \begin{equation}
 \label{perm_def}
 \frac{1}{P}=\int_{0}^{L}{\rm d}z \frac{\exp[\beta G(z)]}{D(z)},
 \end{equation}
where $D(z)$ is the position-dependent diffusion coefficient. The integration boundaries
correspond to the bilayer midplane ($z=0$) and the
membrane extension ($z=L$).

While the computational investment of
atomistic simulations prevents the estimation of the potential of mean
force for many compounds,\cite{swift2013back,lee2016simulation} we
show that coarse-grained simulations can provide an efficient proxy.
Indeed, as a first test of the accuracy of a coarse-grained potential of mean
force in reproducing the permeability of a compound, we computed the $\log P$ value of mannitol in a
DOPC membrane at $T=323 K$, and compared it to reference atomistic
results.\cite{carpenter2014method} We assumed a uniform diffusion
coefficient $D(z)=D\approx0.85\cdot10^{-6}$ $\text{cm}^2/\text{s}$,
by performing a graphical extrapolation of the average value reported
in Ref.~\onlinecite{carpenter2014method}. We obtained $\log P=-6.35$,
in very close agreement with the atomistic value of -6.62.
Although assuming a constant diffusion coefficient represents an
approximation, $D(z)$ was shown to be essentially uniform inside the
membrane environment and for different
compounds.\cite{carpenter2014method} 

\section{Conclusions}
Determing the potential of mean force of the insertion of a small
molecule in a lipid bilayer is highly informative but computationally
demanding.  In this work, we employed high-throughput coarse-grained
molecular dynamics simulations in order to perform a screening across
small molecules of small molecular weight ($\approx 30-160$~Da).  We
establish simple relationships relating the water/octanol partitioning
coefficient to key features of the potential of mean force.  More
specifically, this study allows for a semi-quantitative estimation
given a widely available experimental
measurement---the water/octanol partitioning coefficient, and the coarse-grained
representation in terms of Martini beads. Reference
all-atom simulations found in the literature confirm these
relationships for the range of molecular weights considered.  The
potential of mean force thereby becomes an easily accessible quantity
in drug screening applications, and may be employed in various
contexts, such as predicting permeabilities.

Further, any potential of mean force determined at the coarse-grained
level is informative of not one, but rather a large number of small
molecules.  This unexpected property arises due to the coarse-grained
model's limited number of bead types, which amounts to an identical
representation of chemically-similar molecules.  From a mathematical
perspective, the present mapping from atomistic to coarse-grained
molecular representations is \emph{surjective}---``many-to-one.''  The
size of chemical compound space shrinks with the properties of the
coarse-grained model.  We thus foresee transferable coarse-grained
models to play a role in exploring chemical compound space.

We hope that high-throughput coarse-grained molecular dynamics
simulations will further bring about novel insight into the
thermodynamic properties of small molecules in complex environments.

\section*{ACKNOWLEDGMENTS}

We thank Igor Tetko for providing a copy of the ALOGPS software,
\cite{Tetko2001} and Joseph F.~Rudzinski and Karsten Kreis for
  critical reading of the manuscript.  We acknowledge funding from the
  Emmy Noether program of the Deutsche Forschungsgemeinschaft (DFG).\vspace*{6pt}

\bibliography{biblio} 

\end{document}